\documentclass[iop,apj,appendixfloats]{emulateapj}
\usepackage{apjfonts}

\gdef\msun{$M_{\odot}$}
\lefthead{van Dokkum et al.}
\righthead{}
\slugcomment{}
\begin{document}

\title{Forty-Seven Milky Way-Sized, Extremely Diffuse
Galaxies in the Coma Cluster}

\author{Pieter G.\ van Dokkum\altaffilmark{1}, Roberto Abraham\altaffilmark{2},
Allison Merritt\altaffilmark{1}, Jielai Zhang\altaffilmark{2},
Marla Geha\altaffilmark{1}, and Charlie Conroy\altaffilmark{3}}

\altaffiltext{1}
{Department of Astronomy, Yale University, New Haven, CT 06511, USA}
\altaffiltext{2}
{Department of Astronomy, University of Toronto, Toronto, Canada}
\altaffiltext{3}
{Harvard-Smithsonian Center for Astrophysics, 60 Garden St., Cambridge, MA, USA}

\begin{abstract}
We report the discovery of 47 low surface brightness objects in deep
images of a $3\arcdeg\times{}3\arcdeg$ field centered on the Coma cluster,
obtained with the Dragonfly Telephoto Array. 
The objects
have central
surface brightness $\mu(g,0)$ ranging
from 24 -- 26\,mag\,arcsec$^{-2}$ and effective
radii $r_{\rm eff} = 3\arcsec$ -- $10\arcsec$, as measured from archival Canada
France Hawaii Telescope images.
From their spatial distribution
we infer that most or all of the objects are galaxies in the Coma cluster.
This relatively large distance is surprising as it implies that the
galaxies are very large: with $r_{\rm eff} = 1.5$\,kpc -- 4.6\,kpc their sizes are
similar to those of $L_*$ galaxies even though their median stellar
mass is only $\sim 6\times 10^7$\,\msun. The galaxies are relatively red and
round, with $\langle g-i\rangle = 0.8$ and $\langle b/a\rangle =0.74$.
One of the 47 galaxies is fortuitously covered by a deep
Hubble Space Telescope ACS
observation. The ACS imaging shows a large spheroidal object
with a central surface brightness $\mu_{475} = 25.8$\,mag\,arcsec$^{-2}$,
a Sersic index $n=0.6$, and an effective radius of $7''$,
corresponding to 3.4\,kpc at the distance of Coma.
The galaxy is not resolved into stars, consistent with
expectations for a Coma cluster object.
%To our knowledge such ``ultra-diffuse galaxies'' have not been explicitly
%predicted in galaxy formation models.
We speculate that these ``ultra-diffuse galaxies'' (UDGs)
may have lost their gas supply
at early times, possibly resulting in very high dark matter
fractions.

\end{abstract}


\section{Introduction}

While there have been tremendous advances
in deep, high resolution imaging surveys
over the past decades (e.g., {Scoville} {et~al.} 2007; {Heymans} {et~al.} 2012),
the low surface brightness sky remains relatively unexplored.
The Dragonfly Telephoto Array ({Abraham} \& {van Dokkum} 2014)
was developed with the specific
aim of detecting low surface brightness emission.
It is comprised of eight
Canon 400\,mm f/2.8 II telephoto lenses which all image the same
part of the sky, forming what is effectively a 40\,cm f/1.0 refractor. Four
of the lenses are equipped with an SDSS $g$ filter and four with an SDSS $r$
filter. The lenses are attached to cameras that provide
an instantaneous field of view of $2\fdg 6
\times 1\fdg 9$, sampled with $2\farcs 8$ pixels.

The main science program of Dragonfly is a deep imaging survey of a
sample of nearby
galaxies (see {van Dokkum}, {Abraham}, \&  {Merritt} 2014; {Merritt}, {van Dokkum}, \&  {Abraham} 2014).
%, selected from the Cosmicflows-2 database ( ).
%The first galaxy we observed  was
%M101; we used the Dragonfly data to place constraints on its stellar halo
%({van Dokkum} {et~al.} 2014) and to identify seven low surface brightness galaxies
%that are probably associated with it ({Merritt} {et~al.} 2014).
In the late Spring of 2014 we interrupted this survey to observe
the Coma cluster.
%as we were not aware of any existing  wide-field
%imaging of this field by
%a low surface brightness-optimized telescope.
The
main goal of the Coma observation is to accurately measure
the luminosity and color of the intra-cluster light (ICL).
We are also looking for streams and tidal features, inspired by the
beautiful deep imaging of the Virgo cluster of {Mihos} {et~al.} (2005).

A visual inspection of the reduced images revealed
a large number of
faint, spatially-resolved objects. 
The nature of these objects was
not immediately obvious, as they are not listed in existing
catalogs of faint galaxies in the Coma cluster
(e.g., {Ulmer} {et~al.} 1996; {Adami} {et~al.} 2006).
Furthermore, they seemed to
be too large to be part of the cluster:
typical dwarf galaxies have effective
radii of a few hundred parsecs, which corresponds to much
less than a Dragonfly pixel at the distance of Coma
($D_A = 98$\,Mpc; $D_L=103$\,Mpc).\footnote{Assuming
$cz=7090$\,km\,s$^{-1}$ ({Geller}, {Diaferio}, \& {Kurtz} 1999) and
a Hubble constant of 70\,km\,s$^{-1}$\,Mpc$^{-1}$.}

Expecting that the objects would turn out to be isolated dwarf
galaxies in the foreground of the cluster
we decided to perform a (mostly) objective
selection with the aid of Sloan Digital Sky Survey (SDSS)
and archival Canada France
Hawaii Telescope (CFHT) data,
as described in the next Section. Surprisingly, as we show below, the
objects turn out to be  associated with the Coma cluster after all,
and represent a class of very large, very diffuse
galaxies. Only a handful of similar objects
were known from previous studies
({Impey}, {Bothun}, \& {Malin} 1988; {Bothun}, {Impey}, \& {Malin} 1991;
{Dalcanton} et al.\ 1997).

\section{Identification}

\subsection{Candidates in the Dragonfly Data}

The Coma cluster was observed for 26 hrs, obtained over 25 nights
in the period March -- May 2014. 
The images were reduced
using standard techniques, as described in {van Dokkum} {et~al.} (2014) and
{Merritt} {et~al.} (2014), and projected onto a
common astronometric frame with $2\farcs 0$ pixels. Owing to
large dithers between individual exposures the final $g$ and
$r$ images span $3\fdg 33 \times 3\fdg 33$, centered on 
$\alpha=12^h59^m48\fs{}8$, $\delta=27\degr{}58\arcmin{}51\arcsec$.
The FWHM image quality varies somewhat over the field,
but is typically $\approx 6\arcsec$. The 
limiting depths in the images depend on the spatial scale; on the $10\arcsec$
scales relevant for this paper the $1\sigma$ limits are $\mu(g) \sim
29.3$\,mag\,arcsec$^{-2}$ and $\mu(r)\sim 28.6$\,mag\,arcsec$^{-2}$.

We used SExtractor ({Bertin} \& {Arnouts} 1996) to create an initial catalog of
102,209 objects in the Dragonfly field.
The $g$ and $r$ images were summed to increase the S/N ratio in the
detection image.
For each object two magnitudes were
measured: one based on
the flux in SExtractor's ``AUTO'' aperture, and one in
an aperture with a fixed diameter of $6\arcsec$. Objects were flagged as
possible low surface brightness galaxies if their aperture magnitude
is in the range $20<{\rm AB}<23$  and the difference
between the AUTO and aperture magnitude exceeds 1.8. The latter criterion
rejects isolated stars and compact galaxies, leaving 6624 objects
that are faint and extended at the Dragonfly resolution. 

\begin{figure*}[hbtp]
\epsfxsize=18.3cm
\epsffile[20 20 1550 1080]{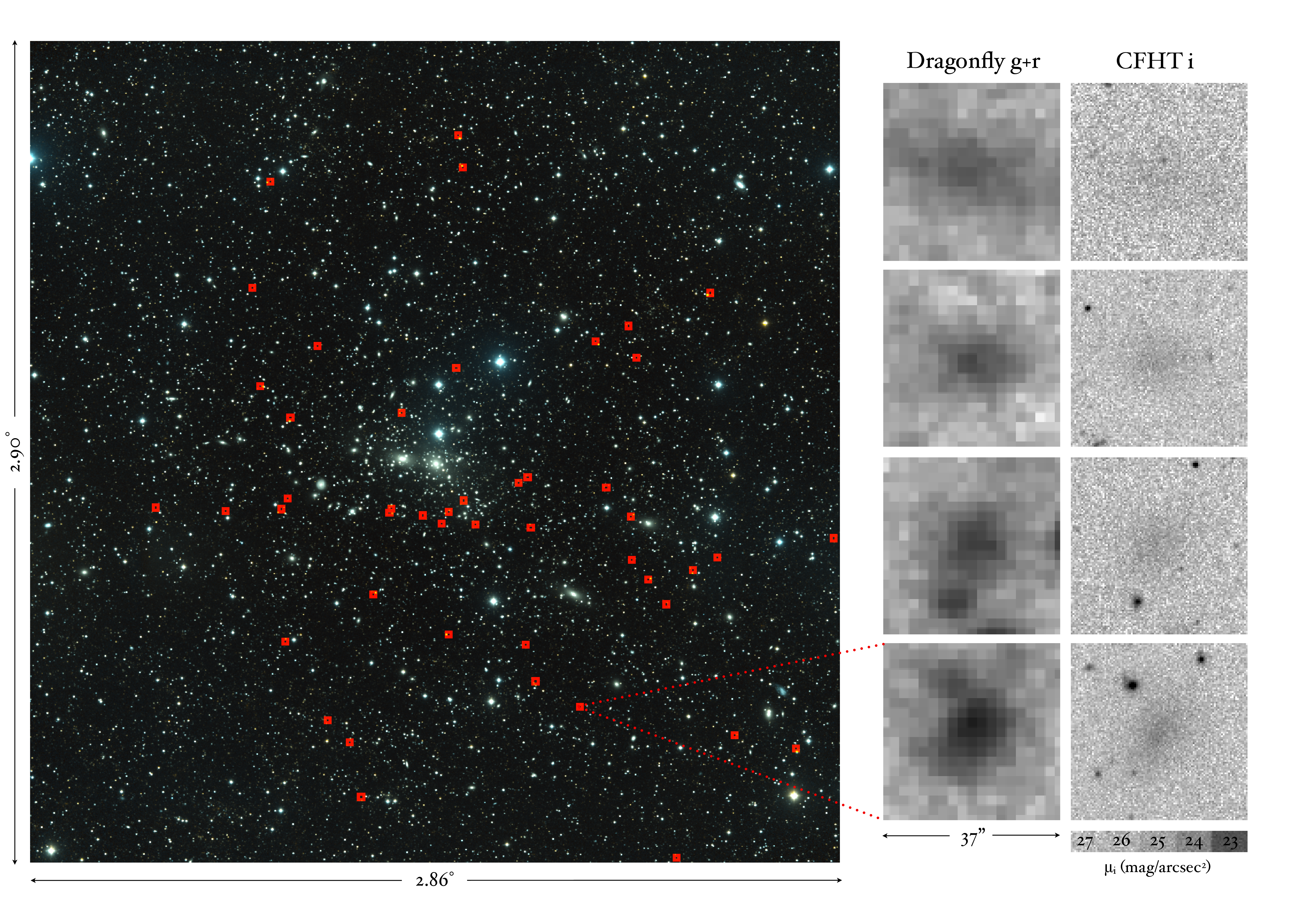}
\caption{\small
Main panel: spatial distribution of the newly discovered galaxies,
projected on a color image of the Coma cluster created from the Dragonfly
$g$ and $r$ images. Only the $2\fdg 86 \times
2 \fdg 90$ area that is covered by CFHT imaging is shown.
Panels at right: typical examples of the galaxies, spanning a range
in brightness.
%They are easily detected but barely resolved in the
%Dragonfly data, and barely detected but easily resolved in the CFHT
%images.
\label{spatial.fig}}
\end{figure*}

\subsection{Rejection Using SDSS and CFHT}

The vast majority of the 6624 objects are not low surface brightness galaxies
but groups of neighboring galaxies, or stars and galaxies,
that are single objects at the Dragonfly resolution.
We removed
most of these by requiring that there is no object in the SDSS catalog
within $4\arcsec$ of the Dragonfly position, leaving 344 candidates.
%In this step we had to
%mask regions around bright stars, as the SDSS is highly incomplete in
%those areas.

The SDSS imaging does not have sufficient depth and spatial
resolution to identify faint groups of galaxies.
We obtained CFHT imaging of the Coma field from the Canadian
Astronomy Data Centre. A $3\arcdeg \times 3\arcdeg$ field was imaged
with a 9-pointing mosaic, in the $g$ and $i$ bands (Head et al.\
2014). Exposure times were
short, at 300\,s per pointing per filter, but the image quality
(FWHM$\,\approx 0\farcs 8$) and sampling ($0\farcs 186$\,pixel$^{-1}$)
are far superior to the Dragonfly and SDSS imaging. We created $37\arcsec \times
37\arcsec$ cutouts of all 344 candidates and used SExtractor to 
identify cases where multiple moderately bright ($i<22.5$)
objects are detected within $7\arcsec$ of the Dragonfly position.
This step left 186 objects which were inspected by eye. Of these, 139
were rejected, with most turning out to be clumps of
multiple objects fainter than the $i=22.5$ limit.

\subsection{A Population of Large, Diffuse Galaxies}

We are left with 47 objects, listed in Table 1,
that are clearly detected in the Dragonfly
imaging, are spatially-extended, are not detected in the SDSS,
and do not resolve into multiple objects in the higher resolution CFHT data.
%Most of the objects that we had noticed when looking at the Dragonfly image
%for the first time are in this sample.
Four typical examples spanning a range of apparent brightness are shown in
Fig.\ \ref{spatial.fig}. The galaxies are clearly detected but barely
resolved in the Dragonfly data, and very faint, fuzzy blobs in the CFHT
data.

We had expected that the objects would be randomly distributed in the
$3\arcdeg \times 3\arcdeg$ field that has both Dragonfly and CFHT coverage,
as their apparent sizes seemed too large for a distance of 100\,Mpc.
However, as shown in Fig.\ \ref{spatial.fig}
they are strongly clustered toward the center of the image.
A Monte Carlo implementation of the
Clark-Evans test gives a probability of 0.04\,\% that the distribution is
spatially-random. Moreover, the apparent East-West
elongation of the distribution is similar to that of confirmed
Coma cluster members (e.g., {Doi} {et~al.} 1995). We conclude that most or
all of the low surface brightness galaxies are, in fact, at the distance
of the Coma cluster and are resolved in the Dragonfly data
because they are intrinsically very large. As we show in \S\,\ref{acs.sec}
this conclusion is supported by Hubble Space Telescope
ACS imaging of one of the galaxies.
%We note that there are no detected galaxies
%within a distance of $11\arcmin$ from the center of the cluster;
%we will return to
%this in \S\,\ref{discussion.sec}.

\section{Properties}

\subsection{Structure}

\begin{figure}[hbtp]
\epsfxsize=8.8cm
\epsffile[15 75 610 685]{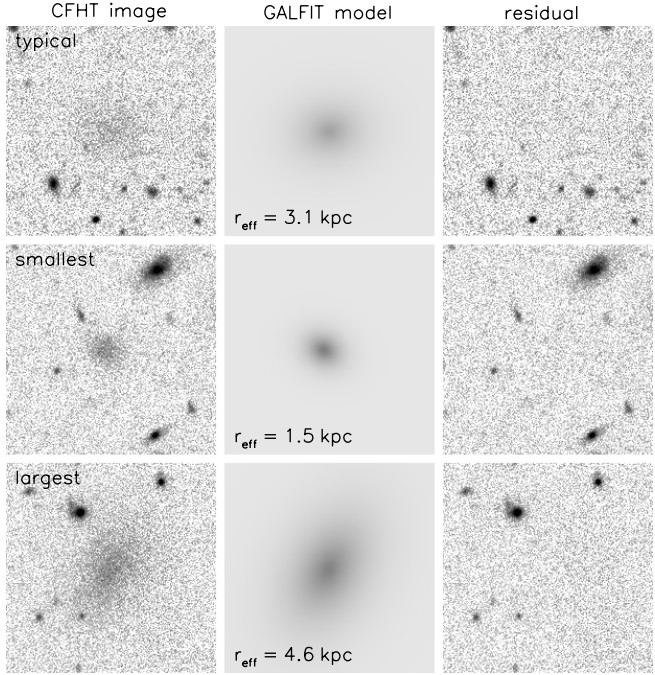}
\caption{\small
Examples of structural parameter fits to the
CFHT data. Each panel spans $37\arcsec \times 37\arcsec$.
The left column shows the summed
$g+i$ images, the middle column shows the best-fitting GALFIT models (with
$n=1$), and
the right column shows the residuals from the fits. The size and
surface brightness of the galaxy in the top (DF1)
row are close to the median of the sample.
The middle row shows the smallest galaxy in the sample (DF43), and the bottom
row shows the largest (DF44).
\label{demo.fig}}
\end{figure}

We used GALFIT ({Peng} {et~al.} 2002) 
to measure structural parameters of the galaxies from the CFHT images.
The fits were performed on the summed $g+i$ images, with neighboring
objects masked. To increase the stability of the fits
the Sersic  index
and sky background were not allowed to vary. All galaxies were fit
three times, with the Sersic index held fixed at $n=0.5$, $n=1$, and $n=1.5$.
The average $\chi^2$ is lowest for $n=1$ (exponential), but for individual
galaxies the three fits are generally equally good. We therefore
use the $n=1$ results for all objects and determine the uncertainties
in the structural parameters of individual galaxies
from the full range of fits.
Three examples of fits are shown in Fig.\ \ref{demo.fig}.
Forty-six galaxies were successfully
fit; the S/N ratio of one object (DF27) is too low for a stable fit.

\begin{figure*}[hbtp]
\epsfxsize=17.5cm
\epsffile[0 144 588 515]{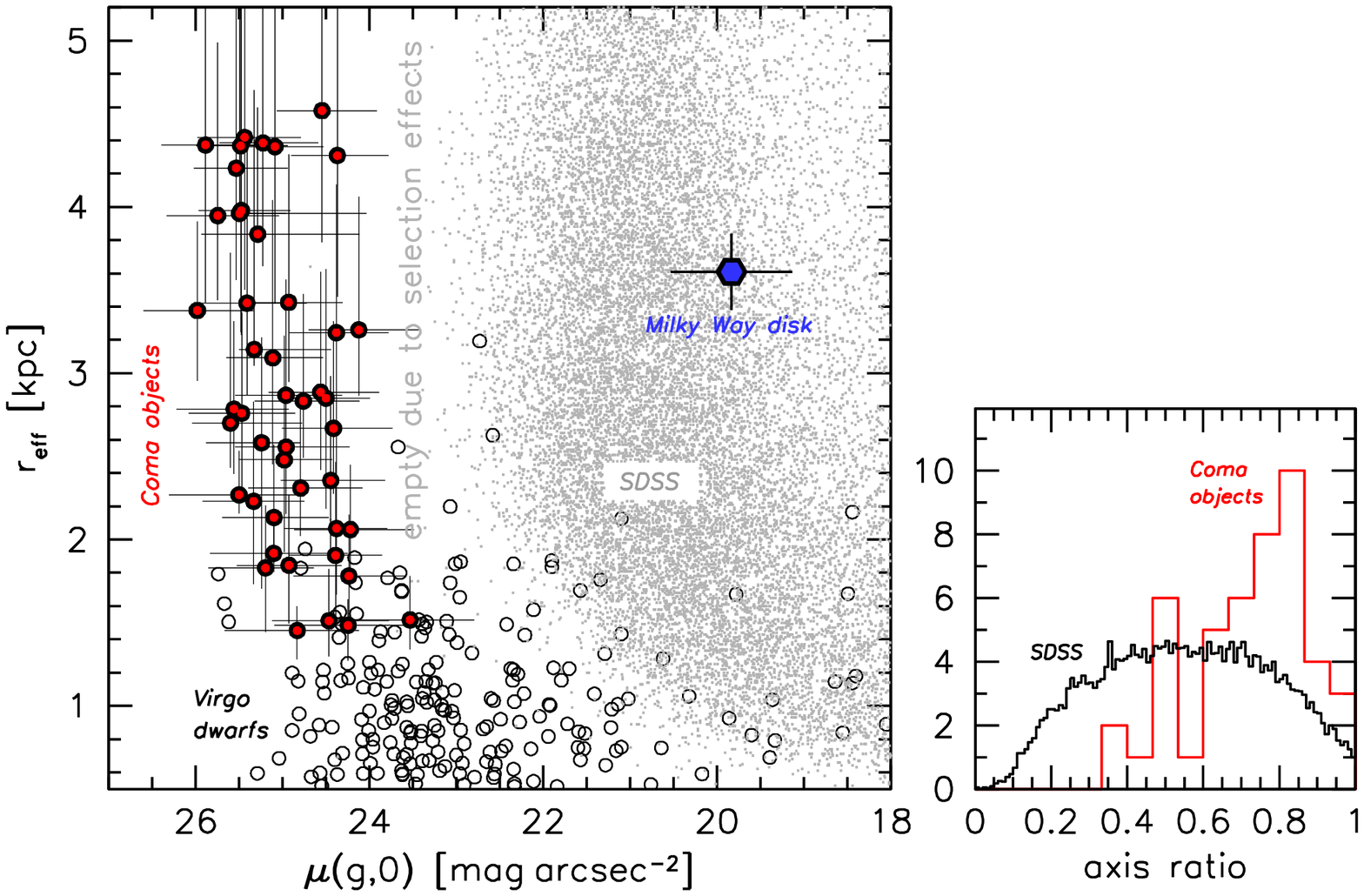}
\caption{\small
Main panel: Location of the newly found galaxies in the effective radius --
central surface brightness plane, compared to galaxies at $0.02<z<0.03$
in the SDSS ({Simard} {et~al.} 2011), early-type
galaxies in the Virgo cluster ({Gavazzi} {et~al.} 2005), and the disk of the
Milky Way ({Bovy} \& {Rix} 2013). 
Right panel: Axis ratio distribution compared to
that of similar-sized SDSS galaxies.
\label{structure.fig}}
\end{figure*}

The distribution of the galaxies in the surface brightness -- size plane
is shown in Fig.\ \ref{structure.fig}, under
the assumption that they are all at the distance of the Coma cluster.
The central surface brightnesses, calculated from the circularized effective
radii and the total fit magnitudes, range from $\mu(g,0)=24-26$\,mag\,arcsec$^{-2}$.
The
effective radii, measured along the major axis, range from 1.5\,kpc to
4.5\,kpc. At fixed surface brightness the newly found galaxies are much larger than
typical dwarf elliptical galaxies in the Virgo cluster
(open circles; {Gavazzi} {et~al.} 2005). 
The median central surface brightness
$\langle \mu(g,0)\rangle =
25.0$\,mag\,arcsec$^{-2}$ ($\approx 25.4$\,mag\,arcsec$^{-2}$
in the $B$ band) and the median effective radius
$\langle r_{\rm eff}\rangle = 2.8$\,kpc.
An interesting point of comparison is the disk
of the Milky Way. {Bovy} \& {Rix} (2013) derive a mass-weighted exponential scale
length of $2.15 \pm 0.14$\,kpc, corresponding to $r_{\rm eff} = 3.6$\,kpc.
Twelve of the newly found objects are larger than this, although for
individual objects the difference is typically not significant.
We note that the gap between SDSS and the Dragonfly
data in Fig.\ 3 is due to the selection limits
of the surveys. The newly found galaxies are
simply the low surface brightness, large size extension of the general
galaxy population, and samples such as that of Thompson \& Gregory (1993)
would fill in the gap.

The axis ratio distribution is shown in
the right panel of Fig.\ \ref{structure.fig}. The galaxies are remarkably
round, with a median axis ratio of $0.74$. We note that
there is no obvious selection effect against inclined disks,
as the galaxies are barely resolved in the Dragonfly data.
Randomly oriented thin disks
would have a uniform $b/a$ distribution, and  this can be ruled out.

\begin{deluxetable*}{ccccccc}
\tablecaption{Positions and Properties}
\tabletypesize{\footnotesize}
\tablewidth{0pt}
\tablehead{\colhead{Id} & \colhead{RA} &\colhead{Dec} &\colhead{$\mu(g,0)$} &
\colhead{$r_{\rm eff}$} & \colhead{$M_g$} & \colhead{$b/a$}\\
\colhead{} & \colhead{(J2000)} & \colhead{(J2000)} &
\colhead{(mag\,arcsec$^{-2}$)}
 & \colhead{(kpc)} & \colhead{(mag)} & \colhead{}}
\startdata
DF1 & 12$^{\rm h}$59$^{\rm m}$14.1$^{\rm s}$ & 29$^{\arcdeg}$07$^{\arcmin}$16$^{\arcsec}$ & $25.1_{-0.5}^{+0.5}$ & $3.1_{-0.6}^{+0.9}$ & $-14.6_{-0.2}^{+0.3}$ & $0.82 \pm 0.03$ \\
DF2 & 12$^{\rm h}$59$^{\rm m}$09.5$^{\rm s}$ & 29$^{\arcdeg}$00$^{\arcmin}$25$^{\arcsec}$ & $24.4_{-0.6}^{+0.6}$ & $2.1_{-0.4}^{+0.6}$ & $-14.3_{-0.2}^{+0.2}$ & $0.71 \pm 0.03$ \\
DF3 & 13$^{\rm h}$02$^{\rm m}$16.5$^{\rm s}$ & 28$^{\arcdeg}$57$^{\arcmin}$17$^{\arcsec}$ & $24.5_{-0.5}^{+0.5}$ & $2.9_{-0.7}^{+0.8}$ & $-14.2_{-0.2}^{+0.3}$ & $0.40 \pm 0.04$ \\
DF4 & 13$^{\rm h}$02$^{\rm m}$33.4$^{\rm s}$ & 28$^{\arcdeg}$34$^{\arcmin}$51$^{\arcsec}$ & $25.7_{-0.6}^{+0.6}$ & $3.9_{-0.5}^{+1.0}$ & $-14.3_{-0.2}^{+0.2}$ & $0.71 \pm 0.03$ \\
DF5 & 12$^{\rm h}$55$^{\rm m}$10.5$^{\rm s}$ & 28$^{\arcdeg}$33$^{\arcmin}$32$^{\arcsec}$ & $24.9_{-0.6}^{+0.6}$ & $1.8_{-0.4}^{+0.4}$ & $-13.5_{-0.2}^{+0.2}$ & $0.71 \pm 0.03$ \\
DF6 & 12$^{\rm h}$56$^{\rm m}$29.7$^{\rm s}$ & 28$^{\arcdeg}$26$^{\arcmin}$40$^{\arcsec}$ & $25.5_{-0.5}^{+0.5}$ & $4.4_{-1.1}^{+1.6}$ & $-14.3_{-0.3}^{+0.4}$ & $0.47 \pm 0.03$ \\
DF7 & 12$^{\rm h}$57$^{\rm m}$01.7$^{\rm s}$ & 28$^{\arcdeg}$23$^{\arcmin}$25$^{\arcsec}$ & $24.4_{-0.5}^{+0.5}$ & $4.3_{-0.8}^{+1.4}$ & $-16.0_{-0.2}^{+0.2}$ & $0.76 \pm 0.03$ \\
DF8 & 13$^{\rm h}$01$^{\rm m}$30.4$^{\rm s}$ & 28$^{\arcdeg}$22$^{\arcmin}$28$^{\arcsec}$ & $25.4_{-0.5}^{+0.5}$ & $4.4_{-0.9}^{+1.5}$ & $-14.9_{-0.3}^{+0.3}$ & $0.73 \pm 0.05$ \\
DF9 & 12$^{\rm h}$56$^{\rm m}$22.8$^{\rm s}$ & 28$^{\arcdeg}$19$^{\arcmin}$53$^{\arcsec}$ & $25.6_{-0.7}^{+0.7}$ & $2.8_{-0.4}^{+0.5}$ & $-14.0_{-0.1}^{+0.1}$ & $0.92 \pm 0.03$ \\
DF10 & 12$^{\rm h}$59$^{\rm m}$16.3$^{\rm s}$ & 28$^{\arcdeg}$17$^{\arcmin}$51$^{\arcsec}$ & $24.4_{-0.6}^{+0.6}$ & $2.4_{-0.4}^{+0.6}$ & $-14.7_{-0.2}^{+0.2}$ & $0.83 \pm 0.03$ \\
DF11 & 13$^{\rm h}$02$^{\rm m}$25.5$^{\rm s}$ & 28$^{\arcdeg}$13$^{\arcmin}$58$^{\arcsec}$ & $24.2_{-0.6}^{+0.6}$ & $2.1_{-0.3}^{+0.4}$ & $-14.8_{-0.1}^{+0.2}$ & $0.98 \pm 0.03$ \\
DF12 & 13$^{\rm h}$00$^{\rm m}$09.1$^{\rm s}$ & 28$^{\arcdeg}$08$^{\arcmin}$27$^{\arcsec}$ & $25.2_{-0.6}^{+0.6}$ & $2.6_{-0.9}^{+0.6}$ & $-14.1_{-0.2}^{+0.5}$ & $0.88 \pm 0.03$ \\
DF13 & 13$^{\rm h}$01$^{\rm m}$56.2$^{\rm s}$ & 28$^{\arcdeg}$07$^{\arcmin}$23$^{\arcsec}$ & $25.3_{-0.6}^{+0.6}$ & $2.2_{-0.5}^{+0.6}$ & $-13.7_{-0.2}^{+0.3}$ & $0.83 \pm 0.03$ \\
DF14 & 12$^{\rm h}$58$^{\rm m}$07.8$^{\rm s}$ & 27$^{\arcdeg}$54$^{\arcmin}$46$^{\arcsec}$ & $25.3_{-0.7}^{+0.7}$ & $3.8_{-0.1}^{+0.8}$ & $-14.4_{-0.1}^{+0.1}$ & $0.51 \pm 0.07$ \\
DF15 & 12$^{\rm h}$58$^{\rm m}$16.3$^{\rm s}$ & 27$^{\arcdeg}$53$^{\arcmin}$29$^{\arcsec}$ & $25.5_{-0.1}^{+0.1}$ & $4.0_{-0.1}^{+5.5}$ & $-14.9_{-0.4}^{+0.1}$ & $0.99 \pm 0.29$ \\
DF16 & 12$^{\rm h}$56$^{\rm m}$52.4$^{\rm s}$ & 27$^{\arcdeg}$52$^{\arcmin}$29$^{\arcsec}$ & $24.8_{-0.8}^{+0.8}$ & $1.5_{-0.2}^{+0.1}$ & $-13.2_{-0.1}^{+0.2}$ & $0.82 \pm 0.10$ \\
DF17 & 13$^{\rm h}$01$^{\rm m}$58.3$^{\rm s}$ & 27$^{\arcdeg}$50$^{\arcmin}$11$^{\arcsec}$ & $25.1_{-0.5}^{+0.5}$ & $4.4_{-0.9}^{+1.5}$ & $-15.2_{-0.2}^{+0.3}$ & $0.71 \pm 0.03$ \\
DF18 & 12$^{\rm h}$59$^{\rm m}$09.3$^{\rm s}$ & 27$^{\arcdeg}$49$^{\arcmin}$48$^{\arcsec}$ & $25.5_{-0.6}^{+0.6}$ & $2.8_{-0.5}^{+0.6}$ & $-13.4_{-0.1}^{+0.2}$ & $0.47 \pm 0.03$ \\
DF19 & 13$^{\rm h}$04$^{\rm m}$05.1$^{\rm s}$ & 27$^{\arcdeg}$48$^{\arcmin}$05$^{\arcsec}$ & $25.9_{-0.5}^{+0.5}$ & $4.4_{-0.9}^{+1.6}$ & $-14.5_{-0.3}^{+0.3}$ & $0.78 \pm 0.03$ \\
DF20 & 13$^{\rm h}$00$^{\rm m}$18.9$^{\rm s}$ & 27$^{\arcdeg}$48$^{\arcmin}$06$^{\arcsec}$ & $25.5_{-0.8}^{+0.8}$ & $2.3_{-0.1}^{+0.3}$ & $-13.0_{-0.1}^{+0.1}$ & $0.53 \pm 0.11$ \\
DF21 & 13$^{\rm h}$02$^{\rm m}$04.1$^{\rm s}$ & 27$^{\arcdeg}$47$^{\arcmin}$55$^{\arcsec}$ & $23.5_{-0.7}^{+0.7}$ & $1.5_{-0.2}^{+0.3}$ & $-14.6_{-0.1}^{+0.2}$ & $0.82 \pm 0.03$ \\
DF22 & 13$^{\rm h}$02$^{\rm m}$57.8$^{\rm s}$ & 27$^{\arcdeg}$47$^{\arcmin}$25$^{\arcsec}$ & $25.1_{-0.6}^{+0.6}$ & $2.1_{-0.3}^{+0.4}$ & $-13.8_{-0.1}^{+0.2}$ & $0.84 \pm 0.03$ \\
DF23 & 12$^{\rm h}$59$^{\rm m}$23.8$^{\rm s}$ & 27$^{\arcdeg}$47$^{\arcmin}$27$^{\arcsec}$ & $24.8_{-0.6}^{+0.6}$ & $2.3_{-0.3}^{+0.5}$ & $-14.3_{-0.2}^{+0.2}$ & $0.89 \pm 0.03$ \\
DF24 & 12$^{\rm h}$56$^{\rm m}$28.9$^{\rm s}$ & 27$^{\arcdeg}$46$^{\arcmin}$19$^{\arcsec}$ & $25.2_{-0.7}^{+0.7}$ & $1.8_{-0.4}^{+0.4}$ & $-12.5_{-0.2}^{+0.2}$ & $0.38 \pm 0.03$ \\
DF25 & 12$^{\rm h}$59$^{\rm m}$48.7$^{\rm s}$ & 27$^{\arcdeg}$46$^{\arcmin}$39$^{\arcsec}$ & $25.2_{-0.5}^{+0.5}$ & $4.4_{-0.7}^{+1.4}$ & $-14.5_{-0.2}^{+0.2}$ & $0.43 \pm 0.03$ \\
DF26 & 13$^{\rm h}$00$^{\rm m}$20.6$^{\rm s}$ & 27$^{\arcdeg}$47$^{\arcmin}$13$^{\arcsec}$ & $24.1_{-0.6}^{+0.6}$ & $3.3_{-0.4}^{+0.8}$ & $-15.4_{-0.2}^{+0.2}$ & $0.63 \pm 0.03$ \\
DF27 & 12$^{\rm h}$58$^{\rm m}$57.3$^{\rm s}$ & 27$^{\arcdeg}$44$^{\arcmin}$39$^{\arcsec}$ & $\gtrsim 26.5$ & \nodata & \nodata & \nodata \\
DF28 & 12$^{\rm h}$59$^{\rm m}$30.4$^{\rm s}$ & 27$^{\arcdeg}$44$^{\arcmin}$50$^{\arcsec}$ & $24.4_{-0.6}^{+0.6}$ & $2.7_{-0.4}^{+0.6}$ & $-14.9_{-0.2}^{+0.2}$ & $0.79 \pm 0.03$ \\
DF29 & 12$^{\rm h}$58$^{\rm m}$05.0$^{\rm s}$ & 27$^{\arcdeg}$43$^{\arcmin}$59$^{\arcsec}$ & $25.3_{-0.2}^{+0.2}$ & $3.1_{-0.1}^{+1.6}$ & $-14.6_{-0.1}^{+0.1}$ & $0.99 \pm 0.13$ \\
DF30 & 12$^{\rm h}$53$^{\rm m}$15.1$^{\rm s}$ & 27$^{\arcdeg}$41$^{\arcmin}$15$^{\arcsec}$ & $24.4_{-0.5}^{+0.5}$ & $3.2_{-0.6}^{+0.9}$ & $-15.2_{-0.2}^{+0.2}$ & $0.70 \pm 0.03$ \\
DF31 & 12$^{\rm h}$55$^{\rm m}$06.2$^{\rm s}$ & 27$^{\arcdeg}$37$^{\arcmin}$27$^{\arcsec}$ & $25.0_{-0.5}^{+0.5}$ & $2.5_{-0.6}^{+0.7}$ & $-14.1_{-0.2}^{+0.3}$ & $0.75 \pm 0.03$ \\
DF32 & 12$^{\rm h}$56$^{\rm m}$28.4$^{\rm s}$ & 27$^{\arcdeg}$37$^{\arcmin}$06$^{\arcsec}$ & $24.8_{-0.6}^{+0.6}$ & $2.8_{-0.3}^{+0.6}$ & $-14.2_{-0.1}^{+0.1}$ & $0.52 \pm 0.03$ \\
DF33 & 12$^{\rm h}$55$^{\rm m}$30.1$^{\rm s}$ & 27$^{\arcdeg}$34$^{\arcmin}$50$^{\arcsec}$ & $25.1_{-0.7}^{+0.7}$ & $1.9_{-0.1}^{+0.2}$ & $-13.4_{-0.1}^{+0.1}$ & $0.69 \pm 0.03$ \\
DF34 & 12$^{\rm h}$56$^{\rm m}$12.9$^{\rm s}$ & 27$^{\arcdeg}$32$^{\arcmin}$52$^{\arcsec}$ & $26.0_{-0.6}^{+0.6}$ & $3.4_{-0.4}^{+0.5}$ & $-13.6_{-0.1}^{+0.1}$ & $0.66 \pm 0.03$ \\
DF35 & 13$^{\rm h}$00$^{\rm m}$35.7$^{\rm s}$ & 27$^{\arcdeg}$29$^{\arcmin}$51$^{\arcsec}$ & $25.6_{-0.4}^{+0.4}$ & $2.7_{-0.3}^{+1.0}$ & $-13.9_{-0.2}^{+0.2}$ & $0.89 \pm 0.09$ \\
DF36 & 12$^{\rm h}$55$^{\rm m}$55.4$^{\rm s}$ & 27$^{\arcdeg}$27$^{\arcmin}$36$^{\arcsec}$ & $25.0_{-0.6}^{+0.6}$ & $2.6_{-0.4}^{+1.0}$ & $-14.3_{-0.4}^{+0.3}$ & $0.80 \pm 0.14$ \\
DF37 & 12$^{\rm h}$59$^{\rm m}$23.6$^{\rm s}$ & 27$^{\arcdeg}$21$^{\arcmin}$22$^{\arcsec}$ & $24.5_{-0.7}^{+0.7}$ & $1.5_{-0.2}^{+0.3}$ & $-13.7_{-0.2}^{+0.2}$ & $0.83 \pm 0.03$ \\
DF38 & 13$^{\rm h}$02$^{\rm m}$00.1$^{\rm s}$ & 27$^{\arcdeg}$19$^{\arcmin}$51$^{\arcsec}$ & $24.2_{-0.6}^{+0.6}$ & $1.8_{-0.3}^{+0.4}$ & $-14.3_{-0.1}^{+0.2}$ & $0.84 \pm 0.03$ \\
DF39 & 12$^{\rm h}$58$^{\rm m}$10.4$^{\rm s}$ & 27$^{\arcdeg}$19$^{\arcmin}$11$^{\arcsec}$ & $25.5_{-0.5}^{+0.5}$ & $4.0_{-0.7}^{+1.3}$ & $-14.7_{-0.2}^{+0.2}$ & $0.77 \pm 0.05$ \\
DF40 & 12$^{\rm h}$58$^{\rm m}$01.1$^{\rm s}$ & 27$^{\arcdeg}$11$^{\arcmin}$26$^{\arcsec}$ & $24.6_{-0.6}^{+0.6}$ & $2.9_{-0.5}^{+0.7}$ & $-14.6_{-0.2}^{+0.2}$ & $0.56 \pm 0.03$ \\
DF41 & 12$^{\rm h}$57$^{\rm m}$19.0$^{\rm s}$ & 27$^{\arcdeg}$05$^{\arcmin}$56$^{\arcsec}$ & $24.9_{-0.5}^{+0.5}$ & $3.4_{-0.5}^{+0.9}$ & $-14.7_{-0.1}^{+0.1}$ & $0.64 \pm 0.03$ \\
DF42 & 13$^{\rm h}$01$^{\rm m}$19.1$^{\rm s}$ & 27$^{\arcdeg}$03$^{\arcmin}$15$^{\arcsec}$ & $25.0_{-0.6}^{+0.6}$ & $2.9_{-0.4}^{+0.6}$ & $-14.1_{-0.1}^{+0.1}$ & $0.52 \pm 0.03$ \\
DF43 & 12$^{\rm h}$54$^{\rm m}$51.4$^{\rm s}$ & 26$^{\arcdeg}$59$^{\arcmin}$46$^{\arcsec}$ & $24.2_{-0.8}^{+0.8}$ & $1.5_{-0.2}^{+0.2}$ & $-13.8_{-0.2}^{+0.2}$ & $0.82 \pm 0.10$ \\
DF44 & 13$^{\rm h}$00$^{\rm m}$58.0$^{\rm s}$ & 26$^{\arcdeg}$58$^{\arcmin}$35$^{\arcsec}$ & $24.5_{-0.5}^{+0.5}$ & $4.6_{-0.8}^{+1.5}$ & $-15.7_{-0.2}^{+0.2}$ & $0.65 \pm 0.03$ \\
DF45 & 12$^{\rm h}$53$^{\rm m}$53.7$^{\rm s}$ & 26$^{\arcdeg}$56$^{\arcmin}$48$^{\arcsec}$ & $24.4_{-0.5}^{+0.5}$ & $1.9_{-0.4}^{+0.6}$ & $-14.2_{-0.2}^{+0.2}$ & $0.80 \pm 0.03$ \\
DF46 & 13$^{\rm h}$00$^{\rm m}$47.3$^{\rm s}$ & 26$^{\arcdeg}$46$^{\arcmin}$59$^{\arcsec}$ & $25.4_{-0.6}^{+0.6}$ & $3.4_{-0.6}^{+1.0}$ & $-14.4_{-0.2}^{+0.2}$ & $0.74 \pm 0.04$ \\
DF47 & 12$^{\rm h}$55$^{\rm m}$48.1$^{\rm s}$ & 26$^{\arcdeg}$33$^{\arcmin}$53$^{\arcsec}$ & $25.5_{-0.5}^{+0.5}$ & $4.2_{-0.7}^{+1.4}$ & $-14.6_{-0.2}^{+0.1}$ & $0.66 \pm 0.04$ 
\enddata
\end{deluxetable*}

\subsection{Stellar Content}

The median absolute $g$ band magnitude $\langle M_g \rangle = -14.3$.
The average color of the galaxies $\langle g-i\rangle = 0.8\pm 0.1$,
as measured from
stacks of the CFHT $g$ and $i$ images.
Their colors are similar to those of the
reddest Milky Way globular clusters (Vanderbeke et al.\ 2014),
and consistent with
an extrapolation of the red sequence
of early-type galaxies in Coma ({Gavazzi} {et~al.} 2010) .
The observed color is consistent with a passively evolving stellar population
with a low metallicity  and/or a relatively young age. For example, 
the {Conroy}, {Gunn}, \& {White} (2009) models predict $g-i=0.8$ for an age of 7 Gyr
and [Fe/H]\,=\,$-1.4$, and for
an age of 4 Gyr and [Fe/H]\,=\,$-0.8$ (see also {Michielsen} {et~al.} 2008).

From the absolute magnitudes
and colors we can estimate the stellar masses of the galaxies. The absolute
magnitudes range from $-16.0 \leq M_g \leq -12.5$;
using Eq.\ 8 in 
{Taylor} {et~al.} (2011) with $g-i=0.8$
we find that the galaxies have stellar masses in the range
$1\times 10^7$\,\msun\ -- $3\times 10^8$\,\msun.
The median stellar mass  $\langle M_{\rm star}\rangle
\sim 6 \times 10^7$\,\msun, and the median
stellar density within the effective radius is
$\sim 5 \times 10^5$\,\msun\,kpc$^{-3}$.

\section{Deep HST/ACS Imaging}
\label{acs.sec}

We searched the HST Archive for serendipitous observations of the newly
found galaxies. Three of the 47 galaxies have been observed by HST. Two of the
observations are short (200\,s -- 300\,s) WFPC2 exposures, which show only
hints of the objects. The third comprises 
8-orbit, multi-band ACS imaging of DF17,
whose properties are close to the median of the sample. The ACS
data include $g_{475}$, $V_{606}$, and $I_{814}$ parallels
to a Cepheid program with the WFC3/UVIS camera
(GO-12476, PI: Cook; {Macri} {et~al.} 2013). The data were obtained from the archive
and reduced using standard techniques (van Dokkum 2001).

\begin{figure*}[hbtp]
\epsfxsize=18cm
\epsffile[0 20 1100 1240]{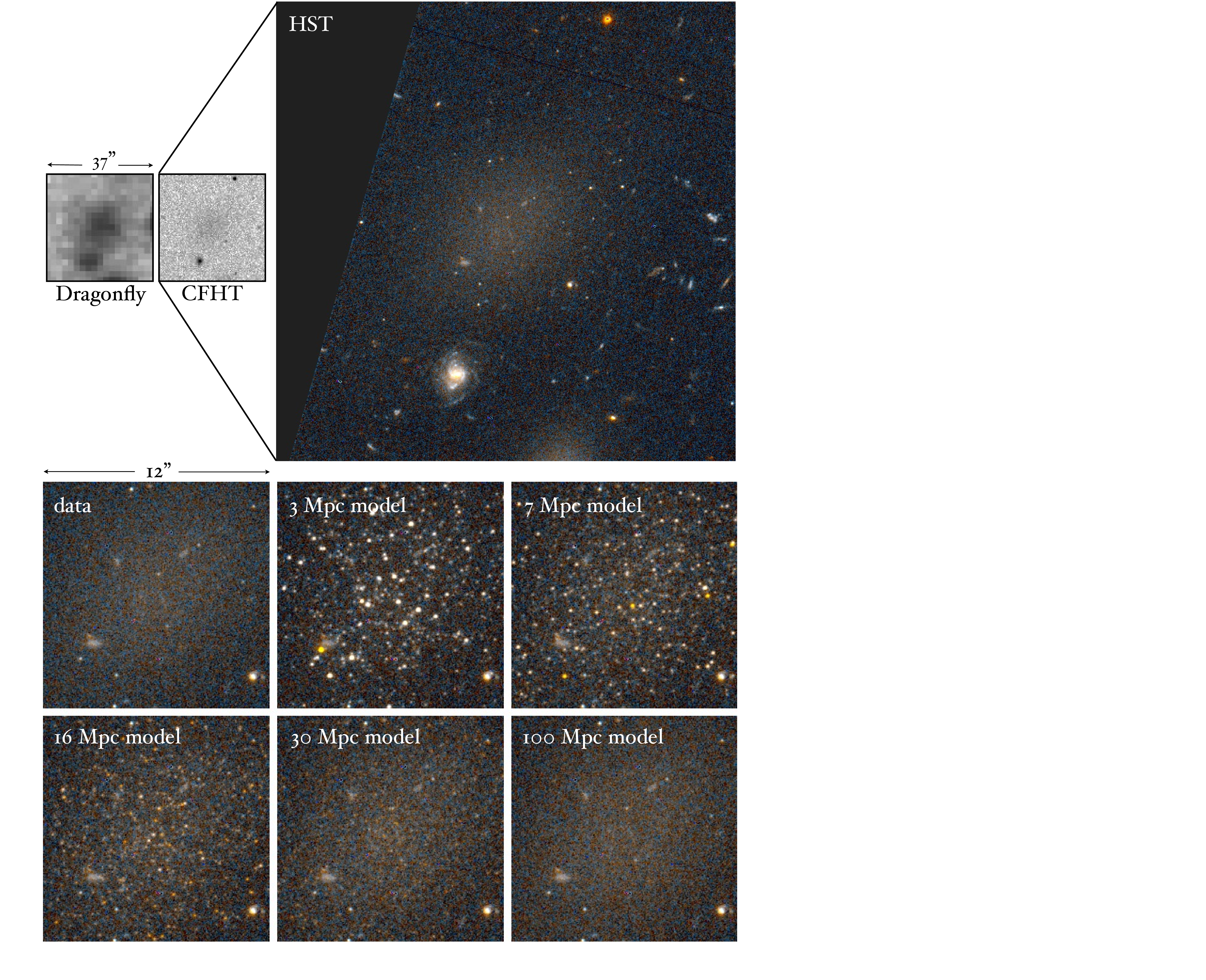}
\caption{\small
One of the galaxies, DF17, has been observed with ACS on HST. The main panel
shows a color image created from the $V_{606}$ and $I_{814}$ ACS
images. The galaxy is smooth, red,
spheroidal, and is not resolved into stars.
The bottom panels show the expected appearance of the galaxy for different
distances (see text). The ACS data are consistent with
the Coma distance of $\approx 100$\,Mpc.
\label{acs.fig}}
\end{figure*}

A color image, created from the $V_{606}$ and $I_{814}$ images, is
shown in Fig.\ \ref{acs.fig}.
DF17 is large and spheroidal and does not have obvious
spiral arms, star forming regions, or tidal features.
We fit the ACS data with a Sersic profile, leaving all parameters
free. The best fitting parameters are
$r_{\rm eff} = 7\farcs 0$,
$n=0.6$, $\mu_{475} = 25.8$, and $b/a=0.71$.
The effective radius, surface brightness, and axis ratio
are in excellent agreement with the
$n=0.5$ fit to the CFHT image.

The fact that the galaxy is not resolved into stars implies a lower
limit to its distance. We created model images of DF17,
following the methodology described in {van Dokkum} \& {Conroy} (2014). Stars were
drawn randomly from a Poisson distribution,
weighted by their expected frequency in a 10\,Gyr old
stellar population with a metallicity [Fe/H]\,=\,$-1.6$.
This stellar population
reproduces the observed $V_{606}-I_{814}$ color 
($V_{606}-I_{814}=0.40$). The models are constrained to reproduce the
observed 2D light distribution of DF17 and its observed total
magnitude of $I_{814}=19.3$, with the distance as the only free parameter.
The model images were convolved with the ACS PSF and
placed in the ACS image, after subtracting the
best-fitting GALFIT model of the galaxy.

The results are shown in the bottom panels of Fig.\ \ref{acs.fig}. Out
to well beyond the Virgo cluster (16 Mpc) the ACS camera easily
resolves individual stars in low surface brightness 
galaxies, as also shown by {Caldwell} (2006).
%Owing to the depth of the
%ACS data we would expect to detect a large population of faint
%stars even at $\sim 30$\,Mpc.
Only at distances $\gtrsim 50$\,Mpc do
the models take on the same smooth appearance as the data, and we conclude
that the ACS observations support the interpretation that the galaxies
are associated with the Coma cluster.
The effective radius of DF17 is then $3.4$\,kpc, almost identical to that of
the disk of the Milky Way.

\section{Discussion}
\label{discussion.sec}

We have identified a significant population of low surface brightness, red,
nearly round objects in a wide field centered on the Coma cluster. 
Based on their spatial distribution and the analysis of one example observed
with ACS we infer that most or all
of the objects are associated with Coma. Their inferred sizes are similar
to those of $L_*$ galaxies and the disk of the Milky Way, even though their
stellar masses are a factor of $\sim 10^3$ lower.

The galaxies do not resemble ``classical'' low surface brightness galaxies (LSBs)
such as those described by, e.g., {van der Hulst} {et~al.} (1993), {Bothun}, {Impey}, \& {McGaugh} (1997),
and {van den Hoek} {et~al.} (2000). Typical LSBs have blue, gas-rich disks, and are
thought to be normal spiral galaxies with a low stellar content and
low star formation rate for their rotation velocity
(see, e.g., {Schombert}, {McGaugh}, \&  {Maciel} 2013, and references therein).
They are also significantly brighter than the objects found in
this paper: the lowest surface brightness object in the compilation
of {Bothun} {et~al.} (1997) has $\mu(0,B) \approx 24.0$\,mag\,arcsec$^{-2}$,
corresponding to $\mu(0,g) \approx 23.6$\,mag\,arcsec$^{-2}$.
Many have bulges; for example, Malin I
has a central surface brightness $\lesssim 16$\,mag\,arcsec$^{-2}$ if
its bulge is taken into account ({Lelli}, {Fraternali}, \& {Sancisi} 2010).

Visually and structurally, the newly found galaxies are more
similar to dwarf spheroidal galaxies such as those found in
the Local Group, around M101, and in the Virgo and Coma clusters than to
classical LSBs: they have similar
Sersic indices, axis ratios, and
surface brightness (e.g., Thompson \& Gregory 1993;
{Geha}, {Guhathakurta}, \& {van der  Marel} 2003; {Gavazzi} {et~al.} 2005; {McConnachie} 2012; {Merritt} {et~al.} 2014; {Toloba} {et~al.} 2014).
However, the term ``dwarf'' is not appropriate for these
large objects.
Dwarf spheroidals have typical sizes of a few hundred pc
(e.g., {McConnachie} 2012; {Lieder} {et~al.} 2012), and
in the Local Group and other nearby groups only a few
have an effective radius exceeding 1\,kpc
(e.g. {Kim} {et~al.} 2011; {McConnachie} 2012; {Chiboucas} {et~al.} 2013; {Merritt} {et~al.} 2014). The largest
known low luminosity Local Group galaxy is And XIX, with a size of
1.6\,kpc
({McConnachie} {et~al.} 2008).
%\footnote{This galaxy is thought to be in the
%process of becoming unbound.}
The Coma objects are much larger, with sizes typical
of $\sim L_*$ spiral and elliptical galaxies (e.g., {Shen} {et~al.} 2003).

\begin{figure}[hbtp]
\epsfxsize=8.6cm
\epsffile{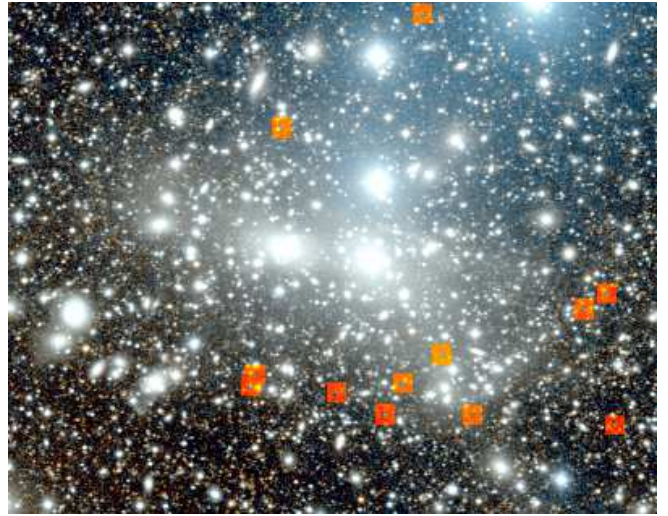}
\caption{\small
Central $0\fdg 89 \times 0\fdg 70$ ($1.6$\,Mpc\,$\times$\,1.2\,Mpc)
of the Dragonfly image shown in Fig.\ 1. 
The newly found galaxies appear to avoid  the region
where  ICL is detected.
\label{coma_cen.fig}}
\end{figure}

The closest analogs to the Coma objects are several very large low
surface brightness objects in the Virgo and Fornax clusters, first
identified by {Impey} {et~al.} (1988).  There are four galaxies in the Impey
et al.\ sample with central surface brightness $\gtrsim
25$\,mag\,arcsec$^{-2}$ and $r_{\rm eff}>2.5$\,kpc; the largest of
these, V1L5 and V4L7, have $r_{\rm eff}=3.7$\,kpc. As the Impey et al.\
survey area is $4\times$ smaller than ours the number
of such galaxies in Virgo and Coma could be similar.
Although the
distances to these particular objects are not confirmed,
{Caldwell} (2006) used HST/ACS imaging to show that at least one
galaxy with a central surface brightness of $\mu(g,0) \approx 27.2$
and an effective radius of 1.5\,kpc is part of the Virgo cluster.
We propose the term ``ultra-diffuse galaxies'', or UDGs, for
galaxies with $r_e\gtrsim 1.5$\,kpc and
$\mu(g,0)\gtrsim 24$\,mag\,arcsec$^{-2}$. We stress that this term
does not imply that
these objects are distinct from the general galaxy population; these
are simply the largest and most diffuse objects in a continuous distribution.

As shown in Fig.\ \ref{coma_cen.fig}  no UDGs are
found in the central regions of the cluster, consistent with
earlier results for slightly brighter diffuse spheroidals in Coma
(Thompson et al.\ 1993).
This could mean that
they are only able to survive at large radii
(see, e.g., {Bothun} {et~al.} 1991; {Gregg} \& {West} 1998; {Martel}, {Barai}, \& {Brito} 2012).
We can estimate what the mass of the galaxies needs
to be to survive a passage within $\sim 300$ kpc of the core of
the cluster, which is where the closest-in UDGs are located.
The criterion for survival is that the total mass $m_{\rm tot}$
within the tidal
radius $r_{\rm tide}=2r_e=6$\,kpc is at least
$m_{\rm tot}> 3 M (r_{\rm tide}/R)^3$,
with $M$ the mass of the cluster within radius $R$.
Using the mass profile of Abell 2667 ({Newman} {et~al.} 2013) as a proxy
for that of Coma, we find $m_{\rm tot}\gtrsim 3\times 10^9$\,\msun,
or a dark matter fraction {\em within the tidal radius} of $\gtrsim 98$\,\%.
We note
that there may be UDGs closer to the cluster core,
as crowding and the ICL limit our
ability to detect them (see {Ulmer} {et~al.} 1996;
{Adami} {et~al.} 2006, 2009). 

It is not clear how UDGs were formed.
It seems unlikely that they are
the product of galaxy harrassment  ({Moore} {et~al.} 1996) or tidal stirring
({Mayer} {et~al.} 2001) of infalling galaxies:
these processes tend to {\em shrink} galaxies,
as the stars at larger radii are less bound than the stars at small radii
(see, e.g., {Mayer} {et~al.} 2001). A likely end-product of cluster-induced tidal effects
are the ultra-compact dwarfs ({Drinkwater} {et~al.} 2003), which
have similar total luminosities and stellar masses as UDGs
but stellar densities that are a factor
of $\sim 10^7$ higher.\footnote{It is remarkable that both classes of objects exist in
clusters at the same time.}
We note, however, that the morphological evolution
of infalling galaxies is difficult to predict, as
it probably depends sensitively on the shape of the inner dark matter
profile (e.g., {Pe{\~n}arrubia} {et~al.} 2010).
%In this context, we cannot exclude the possibility that UDGs are in
%the process of becoming entirely unbound, in which case they
%should have broad tidal tails and distorted morphologies when imaged
%with higher S/N ratio.
%
An intriguing formation scenario is 
that UDGs are ``failed'' $\sim L_*$
galaxies, which lost their gas  after
forming their first generation(s) of stars at high redshift (by
ram pressure stripping or other effects).
If this is the case they
may have very high dark matter fractions, which could also help explain their
survival in the cluster.
Future studies of these objects, as well as
counterparts in other clusters and in the field (see Dalcanton
et al.\ 1997), may shed more
light on these issues.

%these are actually quite bright - in observed brightness and size about
%as bright as m101 dwarfs but much
%more compact. The M101 dwarfs are barely detected even in deep CFHT images
%(vd et al, in preparation) but these eare easy in 5 min. Dragonfly is aimed
%at small distances; at 100 mpc
%CFHT much better.
%Dragonfly's role
%mostly in convenience of a single
%image with large pixels, lack of artifacts, and the fact that we can use
%the cfht data as control so we see
%the objects in two independent datasets.
%sizes 1-1.5 arcmin - much harder to detect. May be much fainter systems around,
%also in Coma.

%clusters at z=0.2 etc - frontier fields - should see many
%but sb dimming

\begin{acknowledgements}
We thank the anonymous referee for an excellent and constructive report.
We also thank the staff at New Mexico Skies for their support
and Nelson Caldwell for comments on the manuscript.
Support from NSF grant AST-1312376 is gratefully acknowledged.
\end{acknowledgements}

%% --------------------------------------------------------------------
%% Mon Oct 27 20:30:31 2014
%%   This file was generated automagically from the files
%%   coma.bbl and coma.tex using
%%     nat2jour.pl
%%   This file should accompany coma-aas.tex.
%% --------------------------------------------------------------------

\end{document}